# Photocatalytic Degradation of Methylene Blue by ZnO/NiFe$_2$O$_4$ Nanoparticles

AdelekeJ.T[1], Theivasanthi.T[2*], Thiruppathi.M[2], Swaminathan.M[2], Akomolafe.T[3], AlabiA.B[3].

1 – Osun State University, Osogbo, Nigeria.
2 - International Research Centre, Kalasalingam Academy of Research and Education (Deemed University), Krishankoil – 626 126, Tamil Nadu, India.
3 – University of Ilorin, Ilorin, Nigeria.
  * Corresponding Author: ttheivasanthi@gmail.com

**Abstract:** *This work reports the successful solid state synthesis of ZnO/NiFe$_2$O$_4$ nanocomposite by calcinations of both green synthesized ZnO nanorods and NiFe$_2$O$_4$ prismatic nanorods at high temperature (850 $^0$C) for ten hours. These nanomaterials were characterized for structural, functional groups and morphological studies using XRD, FTIR and SEM respectively. The photocatalytic activity of the catalysts was assessed through photodegradation of methylene blue (MB) dye. The photocatalytic studies reveal that the ZnO/NiFe$_2$O$_4$ nanocomposite exhibits high degradation efficiency under UV light. The study shows that in the removal of MB solution by ZnO/NiFe$_2$O$_4$ nanoparticles, $^\bullet OH$ and photoproduced holes $h^+$ are the main species. Preparation of ZnO/NiFe$_2$O$_4$ nanocomposite by coupling of NiFe$_2$O$_4$ with green synthesized ZnO in solid state method is the novelty of this report. The synthesized recyclable ZnO/NiFe$_2$O$_4$ nanocomposite has potential applications in the treatment of both domestic and industrial waste water.*

## 1. Introduction

Clean water is the valuable and important thing to the life but pollutants are the major challenge to obtain clean water. Dyes and some organic matters are the major pollutants of water. Pollutants of waste water are mainly come from industries like; textile, photography, printing, painting, leather, agro allied companies – pesticides, insecticides and fertilizers. These water pollutants pose environmental hazards to man, aquatic animals and microorganisms. Generally, it is a threat to the ecosystems due to its toxicity and the carcinogenic effect [1, 2]. Several methods of pollutant removal include; sedimentation, adsorption, chemical precipitation and some biological means but for their limitations. Photocatalysis, an advanced oxidation process, is highly efficient, cost effective; eco-friendly and stable photocatalysts have found their routes through semiconductors in nano forms. TiO$_2$, ZnO, Fe$_2$O$_3$ and CdS, CdO, ZnS have been proved to be suitable photocatalysts for the removal of organic water pollutants [3-5]. Among all these and other photocatalysts, TiO$_2$ is commonly used yet. ZnO has gained more attentions due to its desirable properties – electric, photonic and oxidation resistance, non-toxic, high band gap (Eg = 3.37eV) [6-8].

Doping with transition metals or their oxides had significantly improved the photocatalytic ability of ZnO by impacting shifts in its optical absorption, increasing surface defects and production of surface oxygen vacancies and inhibiting recombination of charge carriers as reported in literatures [9-15]. In recent times, there have been more work on NiFe-based Oxygen evolution reaction (OER) electrocatalysts, from synthesis to catalytic performance in various applications [16,17,*8*]. NiFe oxide doped with TiO$_2$ photocatalysts for heterogeneous photocatalytic water treatment using visible-light-responsive had been reported [18]. Nowadays, for better activity and stability, research is focused on developing advanced NiFe-based materials [19].

Rahmayeni *et al.* reported that the coupling of the ferromagnetic NiFe$_2$O$_4$ with diamagnetic ZnO leads to the appearance of superparamagnetic properties in the synthesized material. This superparamagnetic property of the ZnO/NiFe$_2$O$_4$ composite makes easier the separation of catalyst from the liquid. This nanocomposite can be applied for degradation of dyes under *solar light* [20]. Coupling of ZnO with magnetic materials makes magnetic photocatalyst with recyclable property. It helps for magnetic separation of the photocatalyst from the aqueous solutions after dyes and waste water treatment [21, 22]. Very less works are present in the literatures regarding the photocatalytic performance of ZnO/NiFe$_2$O$_4$ especially on photodegradation of dyes.

Kang *et al.* reported that heat treatment enhances the photo-catalytic reactions and improves the surface area by removing impurities [23]. Hence, in this work, ZnO/NiFe$_2$O$_4$ composite was done in solid state synthesis method by calcinations (with high temperature) of ZnO and NiFe$_2$O$_4$ to prepare an improved photo-catalyst. Synthesis, characterization studies of ZnO/NiFe$_2$O$_4$ nanocomposite and its photocatalytic performances on MB dye removal are being discussed here. This is the first report which discusses about the solid state synthesis of ZnO/NiFe$_2$O$_4$ nanocomposite utilizing green synthesized ZnO nanorods. This recyclable multifunctional photocatalyst has advantage like dye degradation ability under both UV and solar light irradiation. The photodegradation mechanism is shown in Fig. 1.

## 2. Experimental (Materials and Methods)

### 2.1. Precursors
The starting materials for the synthesis of Nickel Ferrite (NiFe$_2$O$_4$) nanorods are FeCl$_3$·6H$_2$O (hydrated iron chloride), NiCl$_2$·6H$_2$O (hydrated nickel chloride), NaOH and (C$_6$H$_9$NO)$_n$ (polyvinylpyrrolidone-PVP). All were of pure analytical grade purchased from Sigma Aldrich. Distilled water was used for all the experiments. For the synthesis of ZnO nanoparticles, the precursors are mango (*Mangifera indica*) leaves extract, anhydrous Zinc acetate [Zn(O$_2$CCH$_3$)$_2$], Sodium hydroxide pellets and ethanol were used.

### 2.2. Synthesis of Nickel ferrite Nanorods (NRs)
FeCl$_3$·6H$_2$O solution (0.4 M concentration) and 0.2 M hydrated nickel chloride [NiCl$_2$·6H$_2$O] solution were prepared separately. Equal volumes of these solutions were mixed and stirred vigorously on magnetic stirrer for 1 h at 80 °C. Then, certain amount of polyvinylpyrrolidone [PVP – (C$_6$H$_9$NO)$_n$] was added in to the solution as a capping agent after which 3.0 M sodium hydroxide (NaOH), was added in drops to the solution kept under constant stirring until brown precipitate is formed. The solution was allowed to cool, filtered and washed repeatedly with de-ionized water to obtain a pH of 7 and dried at 150 °C for 2 hours. The NiFe$_2$O$_4$ nanorods formed was then milled by mortar and pestle to fine powder [5, *11*].

### 2.3 Green Synthesis of ZnO nanoparticles (NPs)
*Mangifera indica* leaves were collected within the compounds of Osun State University, Nigeria. The leaves were air dried and blended into powder. The prepared leave powder 50 g was mixed in 100 ml of distilled water. This mixture was boiled at 70 $^0$C temperature for 30 minutes until the color changed. After cooling, the mixture was filtered, centrifuged and the supernatant solution (leave extract) was collected. Zinc acetate dehydrate [(CH$_3$COO)$_2$Zn.2H$_2$O] solution (0.4 M concentration) was prepared with 50 ml distilled water. The solution was stirred for about 20 minutes. The leave extract (5 ml) was added to the solution under continuous stirring. 2 M sodium hydroxide (NaOH) was added in drops to the mixture to enhance precipitation and to maintain the pH 12. The whole mixture was stirred further for 1 hour. The precipitate was taken out, washed repeatedly with distilled water and then washed with ethanol to remove impurities, after which it was dried in at 60 $^0$C for six hours to yield white ZnO NPs. The synthesized ZnO NPs was calcinated for one hour in air tight oven at 100 $^0$C.

### 2.4 Synthesis of ZnO/NiFe$_2$O$_4$ hybrid nanocomposite
The synthesis route used here was solid state synthesis. NiFe$_2$O$_4$ NRs, ZnO NPs and thiourea were mixed in the ratio of 1:20:10 respectively. The mixture was calcinated in a furnace at 800 $^0$C temperature. After 8 hours of the calcinations ZnO/NiFe$_2$O$_4$ hybrid nanocomposite was obtained. This high temperature level heat treatment enhances the photo-catalytic reactions.

### 2.5 Characterizations Studies
The synthesized ZnO NPs, NiFe$_2$O$_4$ NRs and ZnO/NiFe$_2$O$_4$ nanocomposite were analyzed by using Fourier Transform Infra-Red (FTIR- SHIMADZU - IR TRACER 100) spectroscopy, X-Ray Diffraction (XRD-D8 advance ECO XRD systems with SSD1601D Detector, with Cu-K$\alpha_1$ and K$\alpha_2$ radiation), scanning electron microscopy (SEM- ZEISS-EVO 18 Research).

### 2.6 Photocatalytic activity experiments
The photocatalytic activity of the ZnO/NiFe$_2$O$_4$ NPs was evaluated by photodegradation of an aqueous MB textile dye. The experiment was carried out in a HEBER multilamp Photoreactor (Model – HML-MP88) using four 360 nm UV lamps of 8 W each. The catalytic experiments were carried out with 50 mL solution of MB (5x10$^{-5}$ M concentration) and 20 mg powder of the ZnO/NiFe$_2$O$_4$ NPs catalyst under constant stirring. About 4 mL of the aliquot solution was withdrawn at predetermined time intervals (10 minutes) from the reaction mixture, centrifuged and the decrease in absorbance values were monitored by spectrophotometry. The experiment was repeated for different amounts of catalyst - 40 mg, 50 mg and 75 mg keeping the concentration of the dye solution constant at 5 x 10$^{-5}$ M. A control experiment was carried out under identical experimental condition using commercial MB without catalyst. In addition, degradation of MB dye with 75 mg ZnO/NiFe$_2$O$_4$ catalyst was analyzed, in the absence of irradiation.

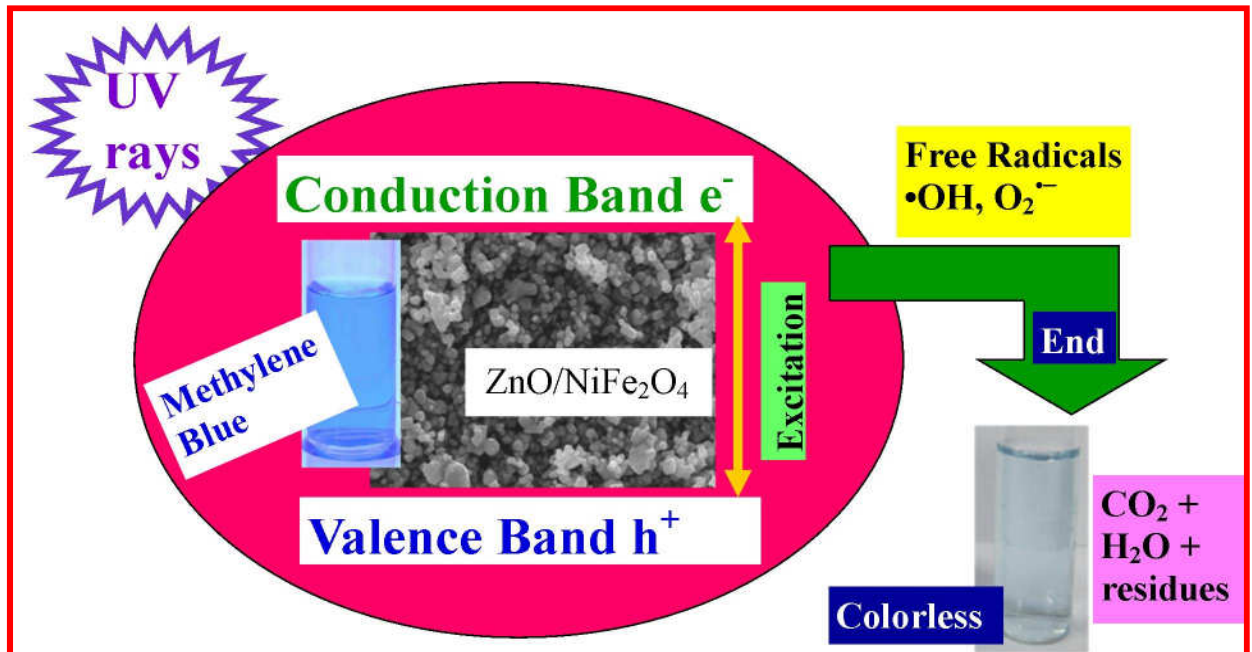

**Fig.1.** Schematic Diagram- Photodegradation of Methylene Blue by photocatalyst ZnO/NiFe$_2$O$_4$

## 3. Results and Discussions
### 3.1 X-ray diffraction analysis

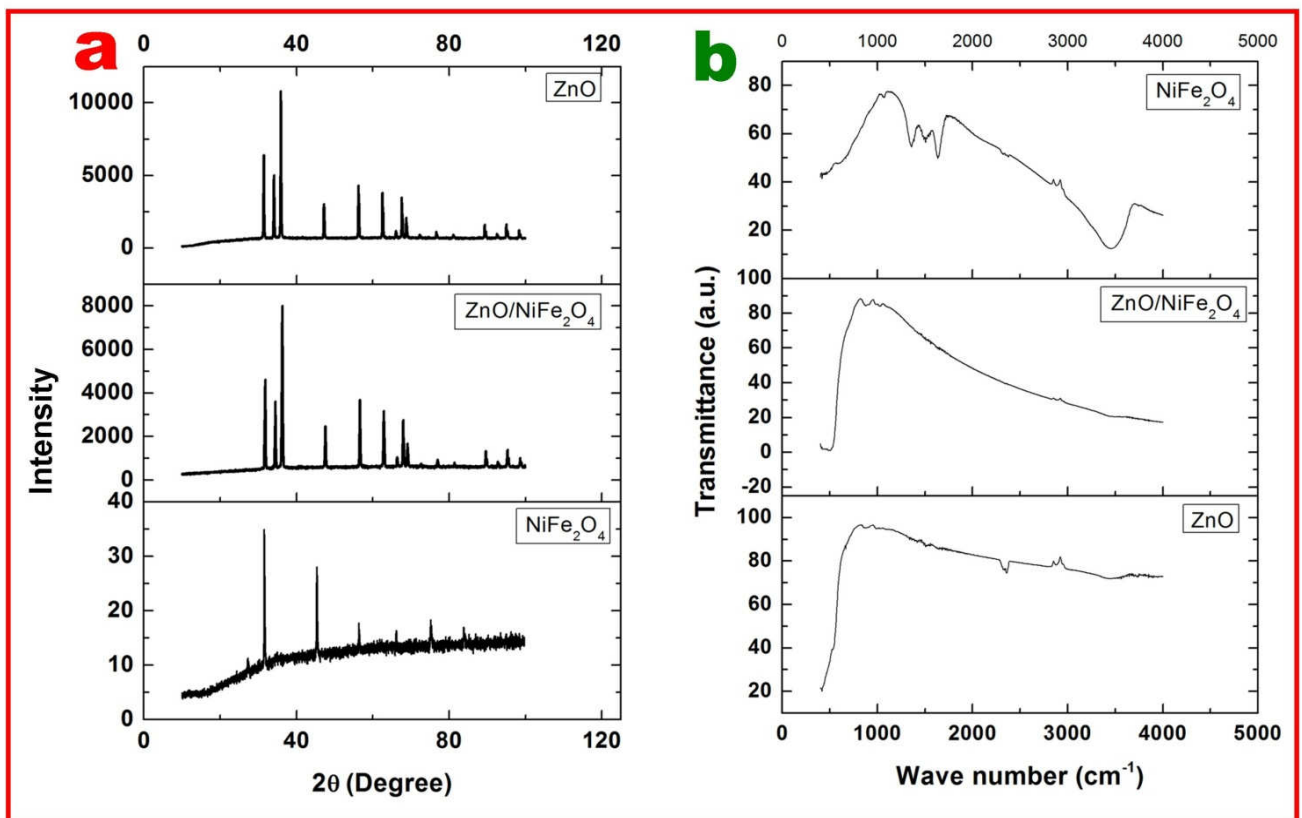

**Fig. 2** (a) XRD patterns (b) FTIR spectra of ZnO NPs, NiFe$_2$O$_4$ NRs and ZnO/NiFe$_2$O$_4$ nanocomposite.

The crystal structures of the green-synthesized ZnO NPs, NiFe$_2$O$_4$ NRs synthesized by co-precipitation method and the hybrid ZnO/NiFe$_2$O$_4$ nanocomposite were studied with XRD (Fig.2.a). The sharp and narrow diffraction peak positions with 2θ values of 31.78, 34.43, 36.27, 47.55, 56.62, 62.88, 66.40, 67.97, 69.11, 72.59, 76.99, 81.41, 89.64 were indexed as (100), (002), (101), (102), (110), (103), (200), (112), (201), (004), (202), (104)

and (203) hkl crystal planes. These peaks confirm formation of a crystalline hexagonal wurtzite structure of the ZnO NPs. These peaks are in agreement with the JCPDS card No. 89 – 0510. Kang *et al.* reported about the wurtzite ZnO [24]. The structure of the prepared ZnO is also agreed with this report.

The $NiFe_2O_4$ NRs diffraction peaks show highly oriented and crystalline structure. The sharp and narrow diffraction peak positions with 2θ values of 21.389, 35.325, 41.701, 43.610, 50.807, 55.768, 63.387, 67.732, were indexed as (111), (220), (311), (222), (400), (331), (422) and (511) hkl crystal planes. These obtained peak intensity profiles are in agreement with the JCPDS card No. 10-0325.

XRD pattern of $ZnO/NiFe_2O_4$ hybrid gives the paeks of both ZnO and $NiFe_2O_4$. The peaks hkl index such as (100), (002), (101), (102), (110), (103), (200), (112), (201), (004), (202), (104) and (203) belong to hexagonal wurtzite ZnO phase while the hkl index such as (220), (222), (400), (331) and (422) correspond to spinel $NiFe_2O_4$ phase. It was observed that there is no extra diffraction peaks of other phases. No obvious peak of impurity implies the purity of the synthesized material. The average grain size of the synthesized ZnO NPs, $NiFe_2O_4$ NRs and $ZnO/NiFe_2O_4$ nanocomposite were calculated as 7.56 nm, 6.67 nm and 7.67 nm respectively, by using Debye-Scherrer's Eq. (1).

$$D = \frac{K\lambda}{\beta Cos\theta} \quad (1)$$

Where D is the average crystallite size (in Å), K is the shape factor, $\lambda$ is the wavelength of X-ray (1.5406 Å) Cu-Kα radiation, β is the full width at half maximum of the diffraction peak, and θ is the Bragg angle [25].

## 3.2 Scanning Electron Microscope (SEM) analysis

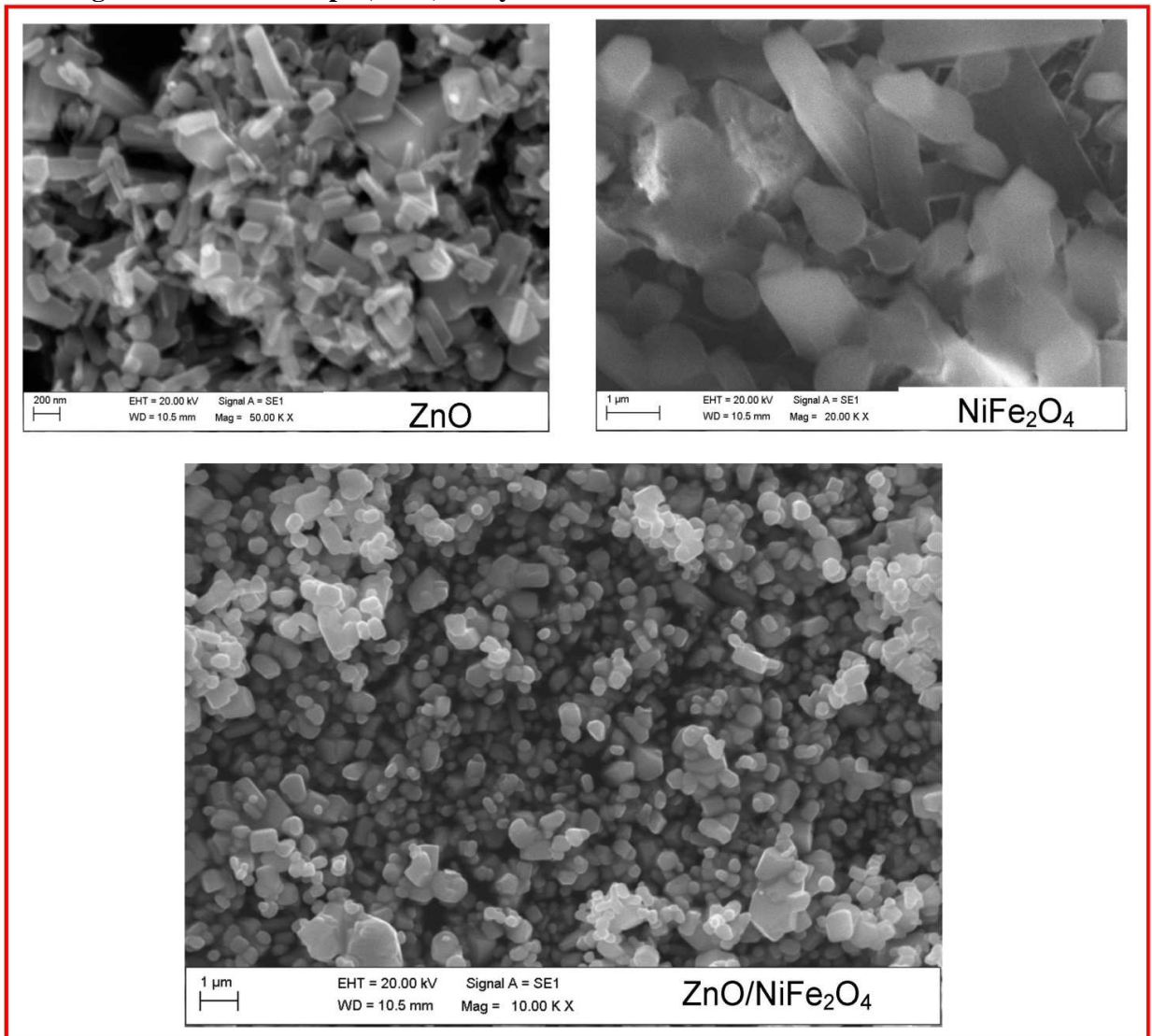

**Fig. 3** SEM images of ZnO NPs, $NiFe_2O_4$ NRs and $ZnO/NiFe_2O_4$ nanocomposite

SEM analysis was used to investigate the structural morphology of the photocatalysts. SEM images of ZnO, $NiFe_2O_4$ and $ZnO/NiFe_2O_4$ are shown in Fig. 3. The SEM image of $NiFe_2O_4$ nanorods shows the rod shape particles with their size range between 10– 20 nm. The green synthesized ZnO nanoparticles' SEM image shows hexagonal shaped nanoparticles. These nanoparticles are in the size range between 10– 25 nm. The SEM image of the $ZnO/NiFe_2O_4$ hybrid shows high degree of agglomeration and different surface morphologies. Yang and Zeng *et al.* reported that these agglomerations and different morphologies may be due to the magnetic attraction between nickel ferrite and zinc oxide layers [26].

### 3.3 Fourier Transform Infrared Spectroscopy (FTIR)
The FTIR spectra of ZnO, $NiFe_2O_4$ and $ZnO/NiFe_2O_4$ are displayed in Figure 2b. The transmittance waveband from 430.0 $cm^{-1}$ to 420.48 $cm^{-1}$ corresponds to the metal-oxygen bonds may be due to ZnO, NiO and $Fe_3O_4$. The stretch around 860 - 900 $cm^{-1}$ implies aromatic ring -1,3- distribution of metal [*19*, 27]. This is in good agreement with Kandasamy *et al.* and Auzans *et al.* [28, 29]. The peaks between 1529.55 - 1500.31 $cm^{-1}$ and 1260 - 1031.92 $cm^{-1}$ indicate O-H in-plane and out-of-plane bonds respectively. The broad peaks around 3700- 3000 $cm^{-1}$ are due to O-H stretch which corresponds to the hydroxyl groups which can be completely removed when the sample is sintered at temperatures $\geq$973 K as reported [30, 31]. In addition to the XRD results, it is observed from the FTIR results that the synthesized materials are ZnO, $NiFe_2O_4$ and $ZnO/NiFe_2O_4$.

### 3.4 Optical Studies
To improve the photo-catalytic performance of photo-catalyst materials doping or coupling with other materials are being done by researchers. Theivasanthi reported that decreasing the band gap (Eg) by doping process prevents the electron-hole pair ($e^-/h^+$) recombination and improves the photocatalytic activities. Such doped $TiO_2$ with a narrow band gap will be applied to absorb both UV and visible light [32]. Reducing the band-gap energy overcomes the problems such as fast recombination rate of electron-hole pairs and light activation under visible light. Non-metal doping into $TiO_2$ particles is an effective method to reduce the band-gap and to enhance catalytic activities [33-34].

$NiFe_2O_4$ nanoparticles absorb visible light due to their small band gap. Short survival of electrons in the conduction band lessens the photo-degradation processes. This drawback restricts the application of this material as a visible light photocatalyst [35]. ZnO strongly absorbs UV light and weakly absorbs visible light. Combining both of these materials improves the visible light absorbance. It will be support for photocatalytic processes under solar light [36].

The approximate band-gap values of ZnO, $NiFe_2O_4$ and $NiFe_2O_4/ZnO$ hybrids are ~3.12 eV, ~1.60 eV and ~1.71 eV respectively. These results suggest that $NiFe_2O_4$ doping enhances the optical absorption property of ZnO nanoparticles [37]. Rahmayeni *et al.* reported the band gap values of ZnO, $NiFe_2O_4$ and $ZnO/NiFe_2O_4$ as 3.12, 1.64 and 2.78 eV respectively [*20*]. Band gap values of Rahmayeni *et al.* report were calculated by Eg = 1240/λ equation [*36*]. Band gap values of ZnO and $NiFe_2O_4$ mentioned in both Zhu *et al.* and Rahmayeni *et al.* reports are well agreed except $ZnO/NiFe_2O_4$ hybrid. However, $ZnO/NiFe_2O_4$ of both reports have band gap values lower than ZnO which lead to better photo-catalytic performance.

Optical studies / photo-catalytic performances of the synthesized bare ZnO and $ZnO/NiFe_2O_4$ nanocomposite have been analyzed under UV light. The analyzed results have been compared. Also, photo degradation of methylene blue (MB) dye solution has been analyzed with two conditions *i.e.* absence of photo-catalysts (self-photo degradation) and absence of UV irradiation. It is observed from the literatures that $ZnO/NiFe_2O_4$ nanocomposite has photocatalytic activity higher than ZnO and $NiFe_2O_4$. Reducing the band-gap of the photocatalyst by doping enhances its performances. The analyzed photo-catalysis results of the synthesized bare ZnO and $ZnO/NiFe_2O_4$ nanocomposite are following.

### 4. Photocatalytic degradation of organic dye

Barzegar *et al.* reported about the photocatalytic degradation of methylene blue (MB) dye by CdS nanoparticles under UV and visible irradiations [38]. Similarly, photocatalytic activity of the $ZnO/NiFe_2O_4$ nanocomposite on the degradation of MB dye was evaluated by the photo decolorization of the dye under UV light. The absorbance spectra of 50 mL solution of MB ($5x10^{-5}$ M concentration) dye were recorded at the irradiation time range between 0 – 70 minutes in the presence of different amount of the photocatalyst ($ZnO/NiFe_2O_4$) powder *i.e.* 20 mg, 40 mg, 50 mg and 75 mg. Fig.4. (a) & (b) show the absorbance spectra for the photocatalyst concentration *i.e.* 50 mg and 75 mg respectively. It can be clearly seen from the photo-decolorization process, the absorbance decreases gradually with increasing of irradiation time. Fig.4. (c) exhibits the linear fitting related to the absorbance spectra of MB dye solution containing 20 mg, 40 mg, 50 mg and 75 mg of ZnO/NiFe$_2$O$_4$ nanocomposite. It is also observed that the photocatalyst with 75 mg concentration is the most efficient.

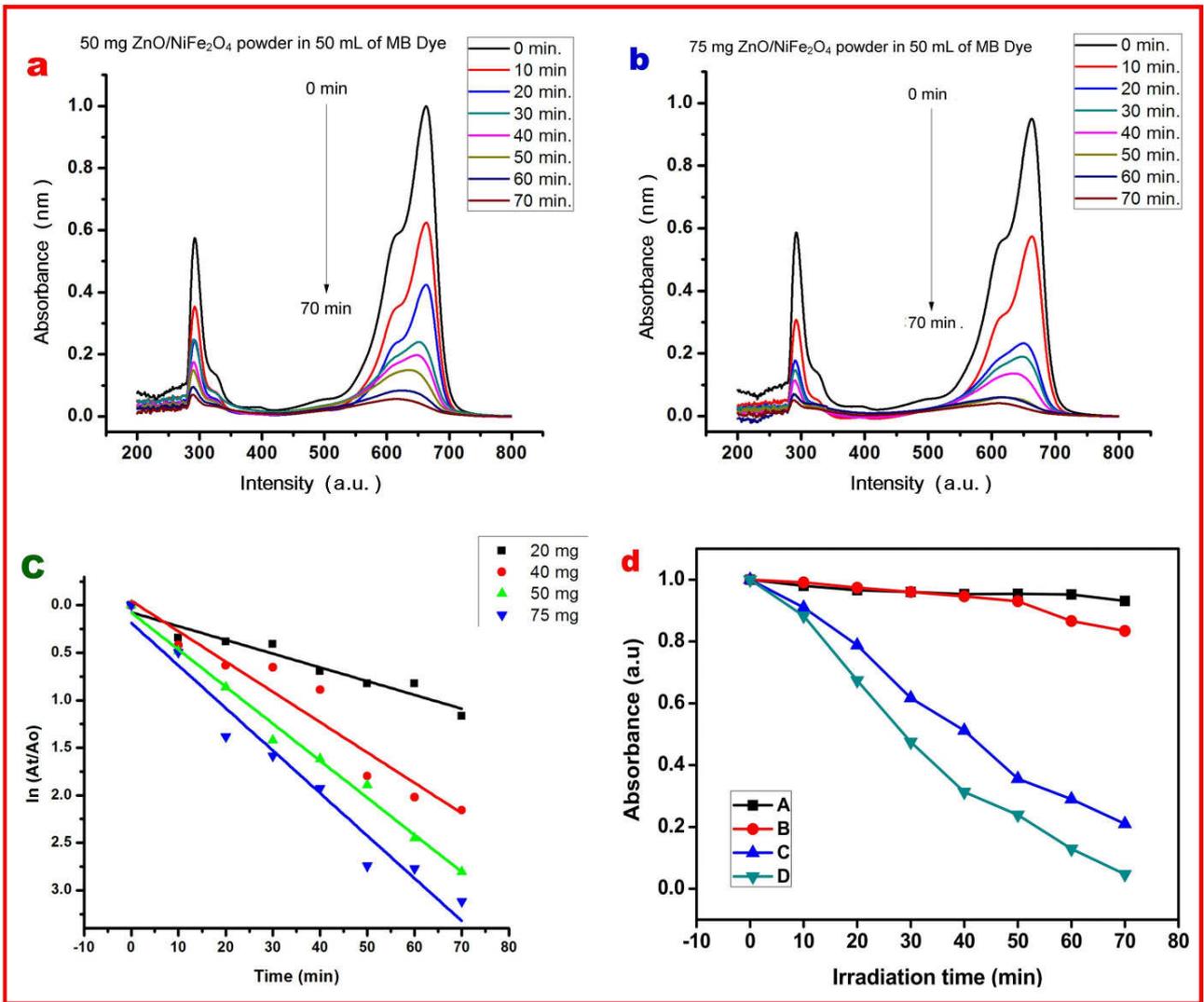

Fig. 4. (a) and (b) Absorbance spectra of 50 mL Methylene Blue [5x10$^{-5}$ M] dye solution containing 50 mg and 75 mg of ZnO/NiFe$_2$O$_4$ nano powder respectively. (c) The linear fitting for the absorbance spectra of MB dye solution containing 20 mg, 40 mg, 50 mg and 75 mg of ZnO/NiFe$_2$O$_4$ nano powder. (d) photo-degradation & photocatalytic performances (A) self-photo degradation of MB dye- with UV irradiation & absence of photo-catalysts (B) MB dye and 75 mg of ZnO/NiFe$_2$O$_4$ -absence of UV irradiation with stirring (C) bare ZnO 75 mg with MB dye (D) ZnO/NiFe$_2$O$_4$ nano powder 75 mg with MB dye.

Fig.4. (d) exhibits the self-photo degradation of MB dye under UV irradiation without ZnO/NiFe$_2$O$_4$. It also shows the ZnO/NiFe$_2$O$_4$ photocatalytic performance in the absence of UV irradiation. When comparing to both of these activities, ZnO/NiFe$_2$O$_4$ has slightly degraded the MB dye in the absence of UV irradiation. In the case of self-photo degradation of MB dye under UV irradiation there is no significant activity in dye degradation. Further, Fig.4 (d) compares the photodegradation performances of the synthesized samples *i.e.* bare ZnO and ZnO/NiFe$_2$O$_4$ on MB dye under UV irradiation. It is observed from this comparison study that ZnO/NiFe$_2$O$_4$ shows more photodegradation performances than bare ZnO.

Fig.4. (c) clearly suggests that the de-colorization efficiency of 75 mg photocatalyst is higher when compared with other concentrations *i.e.* 20 mg, 40 mg and 50 mg.

$$\eta = \frac{C_0 - C}{C_0} \times 100 = \frac{A_0 - A}{A_0} \times 100 \qquad (2)$$

The de-colorization efficiency can be calculated using Eq. (2). Where $C_0$ is the initial concentration of dye and $A_0$ is the corresponding absorbance values. C is the concentration of dye after the irradiation time 't' and its corresponding absorbance value is A [39].

The plots of $\ln(A_0/A)$ against irradiation time drawn for all the samples are shown in Fig. 4. The degradation rate constants (k) were calculated from the slope of the straight line drawn using the linear regression according to the first-order kinetic law $\ln(A_0/A) = kt$ [6, 40]. The calculated k values and the corresponding linear regression coefficient degree ($R^2$) values are shown in Table 1. The values of k and $R^2$ clearly indicate that 75 mg of the $ZnO/NiFe_2O_4$ nanoparticles exhibits better photocatalytic performance. Presence of more reacting ions (species), larger surface area and increased acting sites are improving the photocatalytic performance.

Table 1: Photocatalytic reaction parameters of $ZnO/NiFe_2O_4$ nanoparticles

| Amount of Catalyst (mg) | Rate constant (k), min$^{-1}$ | $R^2$ |
|---|---|---|
| 20.00 | 0.0289 | 0.9315 |
| 40.00 | 0.0617 | 0.9597 |
| 50.00 | 0.0779 | 0.9656 |
| 75.00 | 0.0896 | 0.9897 |

Moreover, the close inter-phase contact coupling of $NiFe_2O_4$ and ZnO nanoparticles in $ZnO/NiFe_2O_4$ nanoparticles should play an important role in enhancing the photoreactivity [41]. Generally, the close coupling of $ZnO/NiFe_2O_4$ nanoparticles results in enhancing the electron transfer rate between interfaces and hindering the recombination of photo-generated electrons and holes [39]. The active species involved in the reaction of photocatalytic degradation of MB dye solution are hydroxyl radicals (•OH), photogenerated holes (h+), and superoxide radical anions ($O_2^{•-}$) [42]. Thus, •OH and h+ play important roles in the photocatalytic degradation of organic pollutants by $ZnO/NiFe_2O_4$ nanoparticles under UV light which is in accordance with the reports from literatures [43, 44].

## 5. Photocatalytic Reaction Mechanism

The first reaction is the adsorption of MB dye molecules on $ZnO/NiFe_2O_4$ nanocomposite shown in Eq.3 which in turn leads to dye-sensitization of ZnO in $ZnO/NiFe_2O_4$ nanocomposite [42, 43]. Under UV light, electrons ($e^-$) in the valence band (VB) of dye-sensitized ZnO and $NiFe_2O_4$ will be excited to the conduction band (CB) while the same amount of holes ($h^+$) in the VB is produced in Eq.4. The CB potential of ZnO is more positive than that of the $NiFe_2O_4$ while the VB potential of $NiFe_2O_4$ is more negative than that of the ZnO [45] Therefore, the photoelectrons ($e^-$) produced in $NiFe_2O_4$ were transferred across the interface of the $ZnO/NiFe_2O_4$ nanocomposite to the surface of ZnO (enumerated in Eq. 5). At the same time, some holes ($h^+$) are transferred quickly from ZnO to $NiFe_2O_4$ due to the more negative VB potential of $NiFe_2O_4$ than that of the ZnO and the close inter-phase contact of $NiFe_2O_4$ and ZnO. The fast migration of $h^+$ and $e^-$ improve the lifetime and transfer of photo-generated charge carriers [46]. The dissolved oxygen ($O_2$) in aqueous solution acting as the electron scavenger react with electrons to yield active free radicals •OH, $O_2^{•-}$, etc. (Eqs. 6 and 7). The separated holes will react with electron donors ($H_2O$) to yield active •OH free radicals (shown in Eq. 8). Subsequently, surface-adsorbed MB molecules were attacked by the generated $h^+$ and other free radicals (•OH, $O_2^{•-}$ etc.), leading to the de-coloration and opening-ring reactions. This is mentioned in equation 9 [46, 47].

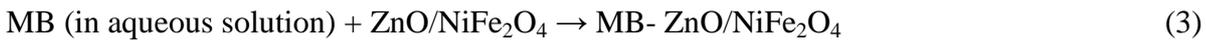

MB (in aqueous solution) + $ZnO/NiFe_2O_4$ → MB- $ZnO/NiFe_2O_4$ (3)

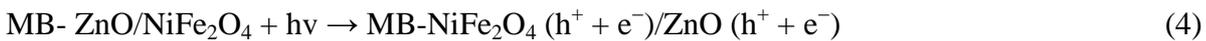

MB- $ZnO/NiFe_2O_4$ + hv → MB-$NiFe_2O_4$ ($h^+ + e^-$)/ZnO ($h^+ + e^-$) (4)

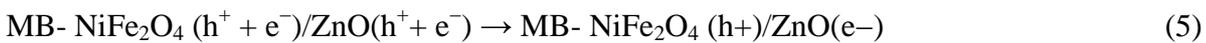

MB- $NiFe_2O_4$ ($h^+ + e^-$)/ZnO($h^+ + e^-$) → MB- $NiFe_2O_4$ (h+)/ZnO(e–) (5)

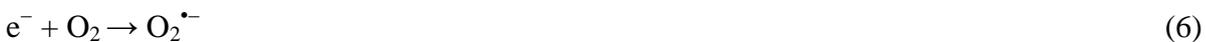

$e^- + O_2 → O_2^{•-}$ (6)

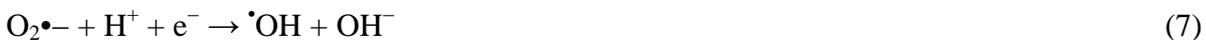

$O_2^{•-} + H^+ + e^- → {}^•OH + OH^-$ (7)

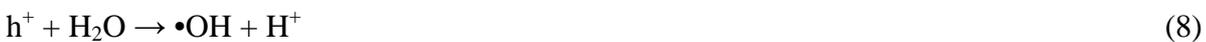

$h^+ + H_2O → •OH + H^+$ (8)

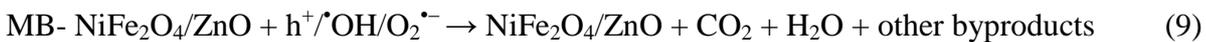

MB- $NiFe_2O_4$/ZnO + $h^+$/$^•OH$/$O_2^{•-}$ → $NiFe_2O_4$/ZnO + $CO_2 + H_2O$ + other byproducts (9)

## Conclusion

ZnO/NiFe$_2$O$_4$ nanocomposite has been successfully synthesized in solid state method by calcination of both green synthesized ZnO and co-precipitated NiFe$_2$O$_4$ at a high temperature of 800 $^0$C for ten hours. This is the first report which discusses about the solid state preparation of ZnO/NiFe$_2$O$_4$ nanocomposite. ZnO/NiFe$_2$O$_4$ nanocomposite exhibits high photocatalytic activity for decolorization of MB solution under UV light. ZnO/NiFe$_2$O$_4$ nanocomposite with 75 mg concentration show highest photocatalytic activity when comparing to other concentrations of the same catalyst. In the removal of MB solution by ZnO/NiFe$_2$O$_4$ nanoparticles, $^\bullet$**OH** and photo-produced holes **h**$^+$ are the main species. This recyclable multifunctional photocatalyst has ability to degrade the organic dyes under UV and solar light irradiations. This study will be a support for the potential applications of synthesized ZnO/NiFe$_2$O$_4$ as photo-catalyst, particularly, in eliminating the hazardous contaminants from both domestic and industrial waste water.

## Acknowledgement

The authors are grateful to the International Research Centre, Kalasalingam Academy of Research and Education (Deemed University), India, Osun State University (Nigeria) and University of Ilorin (Nigeria) for providing supports for this research.